\documentclass{article}

\usepackage{dsfont}
\usepackage{color}
\usepackage{graphicx}
\usepackage{multicol}
\usepackage{multirow,makecell, rotating}
\usepackage[table,xcdraw]{xcolor}
\usepackage{makecell}
\usepackage{acronym}
\usepackage{pifont}
\graphicspath{ {./LaTeX/figures }}
\usepackage[breaklinks]{hyperref}
\usepackage{spconf,amsmath,graphicx}
\usepackage{subcaption}

\usepackage{svg}
\makeatletter
\DeclareRobustCommand*{\escapeus}[1]{%
    \begingroup\@activeus\scantokens{#1 }\endgroup}
\begingroup\lccode`\~=`\_\relax
    \lowercase{\endgroup\def\@activeus{\catcode`\_=\active \let~\_}}
\makeatother

\acrodef{DSP}{Digital Signal Processing}
\acrodef{SSR}{Speech Super Resolution}
\acrodef{LSD}{Log Spectral Distance}

\usepackage{bm}
\usepackage{amssymb,amsmath,graphicx,bbm,color,booktabs}
\usepackage{amsopn}

\newcommand{\vz}{\bm{z}}               

\newcommand{\mynorm}[2]{\| {#1} \|_{#2}}

\newcommand{\twonorm}[1]{\mynorm{#1}{2}}

\renewcommand{\eqref}[1]{Eq.~(\ref{#1})}

\def \x{{\mathbf x}}
\def \y{{\mathbf y}}

\newcommand{\imwav}{\textsc{Im2Wav}}
\newcommand{\imhear}{\textsc{ImageHear}}

\def\x{{\mathbf x}}

\title{I hear your true colors: Image Guided Audio Generation}
\name{Roy Sheffer and Yossi Adi}
\address{School of Computer Science and Engineering \\ The Hebrew University of Jerusalem, Israel}

\begin{document}

\maketitle

\begin{abstract}
We propose \imwav{}, an image guided open-domain audio generation system. Given an input image or a sequence of images, \imwav{} generates a semantically relevant sound. \imwav{} is based on two Transformer language models, that operate over a hierarchical discrete audio representation obtained from a VQ-VAE based model. We first produce a low-level audio representation using a language model. Then, we upsample the audio tokens using an additional language model to generate a high-fidelity audio sample. We use the rich semantics of a pre-trained CLIP (Contrastive Language–Image Pre-training)~\cite{CLIPradford2021learning} model embedding as a visual representation to condition the language model. In addition, to steer the generation process towards the conditioning image, we apply the classifier-free guidance method. Results suggest that \imwav{} significantly outperforms the evaluated baselines in both fidelity and relevance evaluation metrics. Additionally, we provide an ablation study to better assess the impact of each of the method components on overall performance. Lastly, to better evaluate image-to-audio models, we propose an out-of-domain image dataset, denoted as \imhear{}. \imhear{} can be used as a benchmark for evaluating future image-to-audio models. Samples and code can be found under the following \href{https://pages.cs.huji.ac.il/adiyoss-lab/im2wav/}{link}. 
\end{abstract}
\vspace{-0.1cm}
\section{Introduction}
\label{sec:intro}
\vspace{-0.1cm}
Recent advances in neural generative models have challenged the way we create and consume digital content. From image and audio generations~\cite{karras2019style, oord2016wavenet} to the recently proposed textually guided generative methods~\cite{dalle2, nichol2021glide,gafni2022make}, these models have shown remarkable results.

Large-scale datasets of text-image pairs automatically obtained from the internet~\cite{schuhmann2022laion} were one of the main factors enabling recent breakthroughs in such models~\cite{dalle2, nichol2021glide}. However, replicating this success for audio is limited, as a similarly sized text-audio pairs dataset cannot be easily collected. For comparison, DALL-E 2 text-to-image model was trained on $\sim$650M text-image pairs~\cite{dalle2}, while the audio equivalent, \textsc{AudioGen} model~\cite{kreuk2022audiogen} was trained on $\sim$3M text-audio pairs. Contrary to text-audio pairs, videos that can be easily obtained from the web naturally contain image-audio pairs~\cite{Vggsound}. This makes the use of video data appealing for designing a conditional audio generation model.

Generating open-domain visually guided audio is a challenging task. Most prior attempts to solve this task have used a class-aware approach. Chen et al.~\cite{chen2018visually} proposed learning the delta from a per-class average spectrogram representation to an audio instance given input images. Next, the authors in~\cite{zhou2018visual, chen2020regnet} proposed training a model for each class independently. Although these methods provide high-quality generations, they are limited in their generalization ability to unseen classes and require labeled data. Lastly, the current state-of-the-art is the SpecVQGAN model proposed by~\cite{SpecVQGAN_Iashin_2021}. SpecVQGAN is based on a single model capable of generating a diverse set of sounds conditioned on visual inputs from multiple classes without a pre-determined class set. It is conditioned on image representations obtained from a pre-trained image classifier and generates a mel-spectrogram. Then, the generated mel-spectrogram is converted to the time domain using a neural vocoder~\cite{kumar2019melgan}.

\begin{figure}[t!]
\centering
\escapeus{\includegraphics[width=\columnwidth]{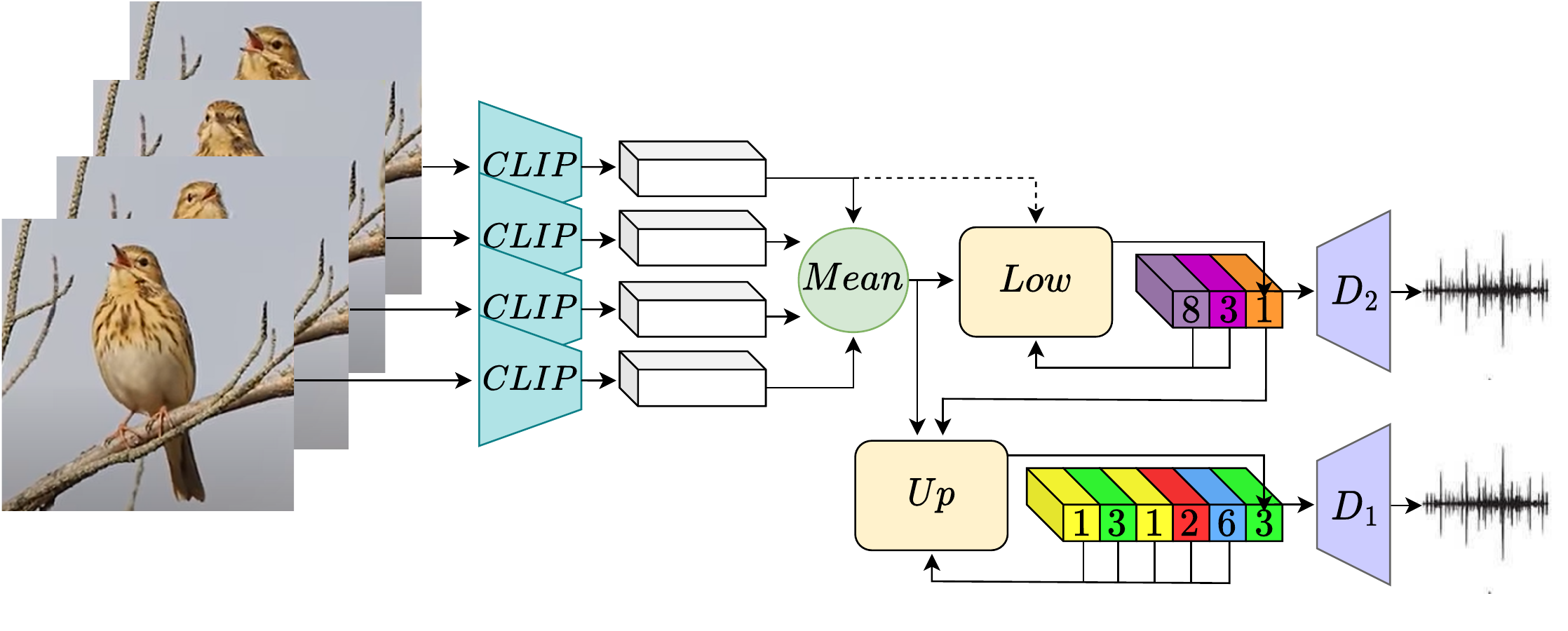}}
\caption{A high-level description of the \imwav{} architecture. Given an image sequence, CLIP features are extracted from each image and used as a condition for an autoregressive audio tokens generation model. The \textsc{Low} level tokens are then upsampled to higher resolution \textsc{Up} level tokens using an additional autoregressive model. Finally, both token sequences are decoded to a time-domain audio signal.\label{fig:Transformerarch}}
\vspace{-0.2cm}
\end{figure}

In this work, we follow such a label-free approach of generating general audio from natural images. Inspired by ~\cite{kreuk2022audiogen, dhariwal2020jukebox}, we propose \imwav{}, a Transformer-based audio Language Model (LM) conditioned on image representation. Given an input image sequence, \imwav{} generates an audio sample that highly correlates with the appeared objects in the image sequence. \imwav{} consists of two main stages. The first encodes raw audio to a discrete sequence of tokens using a hierarchical VQ-VAE model. In the second stage, we optimize an autoregressive Transformer language model that operates on the discrete audio tokens obtained from the first stage. The language model is conditioned on visual representations obtained from a pre-trained CLIP model~\cite{CLIPradford2021learning}. Our proposed model can be conditioned on either a single image or a sequence of images (i.e., video). Additionally, we apply the classifier-free guidance method~\cite{ho2021classifier} to better achieve image adherence in the generation process. 

We empirically show that the proposed method significantly outperforms the evaluated baselines across a diverse set of metrics. We additionally provide a label distribution analysis of the generated audio, together with an ablation study, to better assess the effect of each component of the proposed system. A visual description of the proposed system can be seen in Figure~\ref{fig:Transformerarch}.  


\vspace{-0.2cm}
\section{Method}
\label{sec:method}
\vspace{-0.1cm}
Inspired by previous work~\cite{kreuk2022audiogen}, the proposed system has three main components: (i) an audio encoder-decoder which encodes and decodes audio to and from a discrete representation; (ii) a pre-trained image encoder; and (iii) an audio language model which operates over the discrete audio tokens. 

Formally, we are given an audio-images dataset $\{\mathbf{\x}^{i}, \mathbf{\y}^{i}\}_{i=1}^N$ where $\mathbf{\x}^{i}$ is an audio sample and $\mathbf{\y}^{i}$ is its corresponding sequence of images. Our goal is to learn a function which generates an audio file given a sequence of images (i.e., video). To do so, we first train a VQ-VAE model which allows to represent audio as a discrete sequence of tokens sampled at lower frequency. Then, we train a Transformer-decoder language model over the discrete tokens conditioned on an image representation. During inference time, we sample from the Transformer-decoder to generate a new set of tokens semantically relevant to the input image sequence. We refer to the audio as a $T$ long sequence $\mathbf{\x}^{i} = \langle x^{i}_{t}\rangle_{t=1}^T \in A$.

\subsection{Audio Encoder \& Decoder.}
We use a one dimensional hierarchical VQ-VAE architecture, similar to the one proposed by~\cite{VQ-VAE-2} to encode audio into a discrete space $Z$. The VQ-VAE consists of an encoder $E: A \mapsto H$ which encodes $\mathbf{\x}\in A$ into a sequence of latent vectors $\mathbf{h} = \langle {\mathbf{h}}_{s}\rangle_{s=1}^{S} \in H$. A bottleneck $Q: H \mapsto Z$ that quantizes $\mathbf{h}$ by mapping each $\mathbf{h}_{s}$ to its nearest vector $\mathbf{c}_{j}$, from a codebook $\mathbf{C} = \{\mathbf{c}_{k}\}_{k=1}^{K}$, resulting in a discrete sequence 
$\vz$ = $\langle z_{s}\rangle_{s=1}^{S} \in Z, z_{s}\in 1,\ldots, K$. Then, a decoder $D: Z \mapsto A$ employs the codebook look-up table and decodes the latent vectors back to a time domain signal. The VQ-VAE is trained with the VQ-VAE loss functions as described in~\cite{VQ-VAE-2}, together with a STFT spectral loss similar to the one proposed by~\cite{dhariwal2020jukebox}.

As in \cite{VQ-VAE-2, dhariwal2020jukebox}, we train a single encoder and decoder but break up the latent sequence $\mathbf{h}$ into a multi-level representation $\mathbf{h}=[\langle \mathbf{h}^{(1)}_{s}\rangle_{s=1}^{S^{(1)}},\ldots, \langle \mathbf{h}^{(L)}_{s}\rangle_{s=1}^{S^{(L)}}]$ with decreasing sequence lengths $S^{(l+1)} < S^{(l)}$, each learning its own codebook $\mathbf{C}^{(l)}$. 
\subsection{Image Encoder.}
\label{sec:clip}
We use a pre-trained CLIP \cite{CLIPradford2021learning} model as our image encoder. The CLIP model was trained to maximize the similarity between corresponding text and image inputs. The premise behind using CLIP embedding instead of a pre-trained image classification model, as done in prior work~\cite{SpecVQGAN_Iashin_2021}, is to leverage the semantic information obtained from multi-modal learning. We hypothesize that similar to bilinguals showing advantages over monolinguals when acquiring an additional language~\cite{bilingualism}, modeling an additional modality (audio tokens) may be easier when considering representations from encoders that were optimized with multi-modal data.
In order to convert a sequence of images to a single vector representation denoted as $\tilde{\mathbf{\y}}^{i}$, we average the extracted image features $\langle \mathbf{f}^{i}_{m}\rangle_{m=1}^{\#frames}$ along the time axis and pass it through three MLP layers with ReLU activations.
\subsection{Sequence Modeling.} 
We train two auto-regressive models, denoted as \textsc{Low} and \textsc{Up} in order to learn a prior $p(\vz)$ over the discrete space at two different time resolutions. We utilize an auto-regressive Sparse Transformer Decoder \cite{sparse, allyouneed, dhariwal2020jukebox} causal language model that predicts future audio tokens, conditioned on $\tilde{\mathbf{\y}}^{i}$. 
At every time step, we condition the \textsc{Low} model on the image representation corresponding to the same temporal position $\mathbf{f}^{i}_{\hat{m}}$, together with a positional embedding of the current token offset. For the \textsc{Up} model, we follow a similar setup as in~\cite{dhariwal2020jukebox}, and employ the same Transformer architecture to reconstruct the higher resolution \textsc{Up} level tokens, conditioned on the corresponding \textsc{Low} level generated tokens together with $\tilde{\mathbf{\y}}^{i}$. 

Thus, our objective can be described as maximum-likelihood estimation over the discrete spaces learned by the VQ-VAE as follows,
\begin{equation}
    \label{eq:Transformer}
    \begin{aligned}
    &\max_{\theta_{Low}}\sum_{i=1}^{N} \sum_{s=1}^{S^{(2)}} \log p_{\theta_{Low}}(z^{i}_{s} \mid \tilde{\mathbf{\y}}^{i}, \mathbf{f}^{i}_{\hat{m}}, z^{i}_{1}, \ldots, z^{i}_{s-1}),\\
    &\max_{\theta_{Up}}\sum_{i=1}^{N} \sum_{s=1}^{S^{(1)}} \log p_{\theta_{Up}}(u^{i}_{s} \mid \tilde{\mathbf{\y}}^{i}, \hat{z}^{i}_{s}, u^{i}_{1}, \ldots, u^{i}_{s-1}),
    \end{aligned}
\end{equation}
where $\hat{z}^{i}_{s}, \mathbf{f}^{i}_{\hat{m}}$ are the \textsc{Low} level token and image representation which are mapped to the same temporal position in the input space as the $u^{i}_{s}, z^{i}_{s}$ respectively, and $\theta_{Low}$ and $\theta_{Up}$ are the parameters of the \textsc{Low} and \textsc{Up} auto-regressive models, respectively. Intuitively, as the \textsc{Low} level encodes longer audio per token, it abstractly determines the semantic foundations of the generated audio, while the \textsc{Up} level completes the fine details in higher resolution. Notice, we assume evenly spaced frames to support an arbitrary number of images as input. 

\subsection{Classifier Free Guidance.} 
To further improve the generation performance, and steer the generation process towards the input images, we apply the Classifier-Free Guidance (CFG) method. It was recently shown by the authors in~\cite{ho2021classifier, nichol2021glide} that using the CFG method is an effective mechanism for controlling the trade-off between sample quality and diversity. We follow the same setup as in~\cite{kreuk2022audiogen} in which during training for each sample in the batch with probability $p=0.5$ we replace $\mathbf{\y}^{i} = \langle \mathbf{f}^{i}_{m}\rangle_{m=1}^{\#frames}$ with a learned-null embedding of the same size  $\mathbf{\y}^{\emptyset}=\langle \mathbf{f}^{\emptyset}\rangle_{m=1}^{\#frames}$.
We empirically found that applying CFG to the \textsc{Low} model only is enough to greatly improve the performance. During inference we produce token distributions with and without visual conditioning, and we sample from the following,
\begin{equation}
    \label{eq:cfg}
    \begin{aligned}
    &\log p_{\theta_{Low}}(z^{i}_{s}) = \lambda_{\mathbf{\y}^{\emptyset}} + \eta \cdot (\lambda_{\mathbf{\y}^{i}} - \lambda_{\mathbf{\y}^{\emptyset}}), \\
    &\lambda_{\mathbf{\y}^{i}} = \log p_{\theta_{Low}}(z^{i}_{s} \mid \tilde{\mathbf{\y}}^{i}, \mathbf{f}^{i}_{\hat{m}}, z^{i}_{1}, \ldots, z^{i}_{s-1}), \\
    &\lambda_{\mathbf{\y}^{\emptyset}} = \log p_{\theta_{Low}}(z^{i}_{s} \mid \tilde{\mathbf{\y}}^{\emptyset}, \mathbf{f}^{\emptyset}, z^{i}_{1}, \ldots, z^{i}_{s-1}), \\
    \end{aligned}
\end{equation}
where $\eta\geq1$ is the guidance scale that determines the trade-off between diversity and quality of the generated audio characteristics. We use $\eta=3$ which showed to perform the best in prior works in the fields of text-to-image generation~\cite{nichol2021glide} and text-to-audio generation~\cite{kreuk2022audiogen}.
\begin{figure*}[t!]
     \centering
     \begin{subfigure}[b]{0.49\textwidth}
         \centering
         \includegraphics[width=\columnwidth]{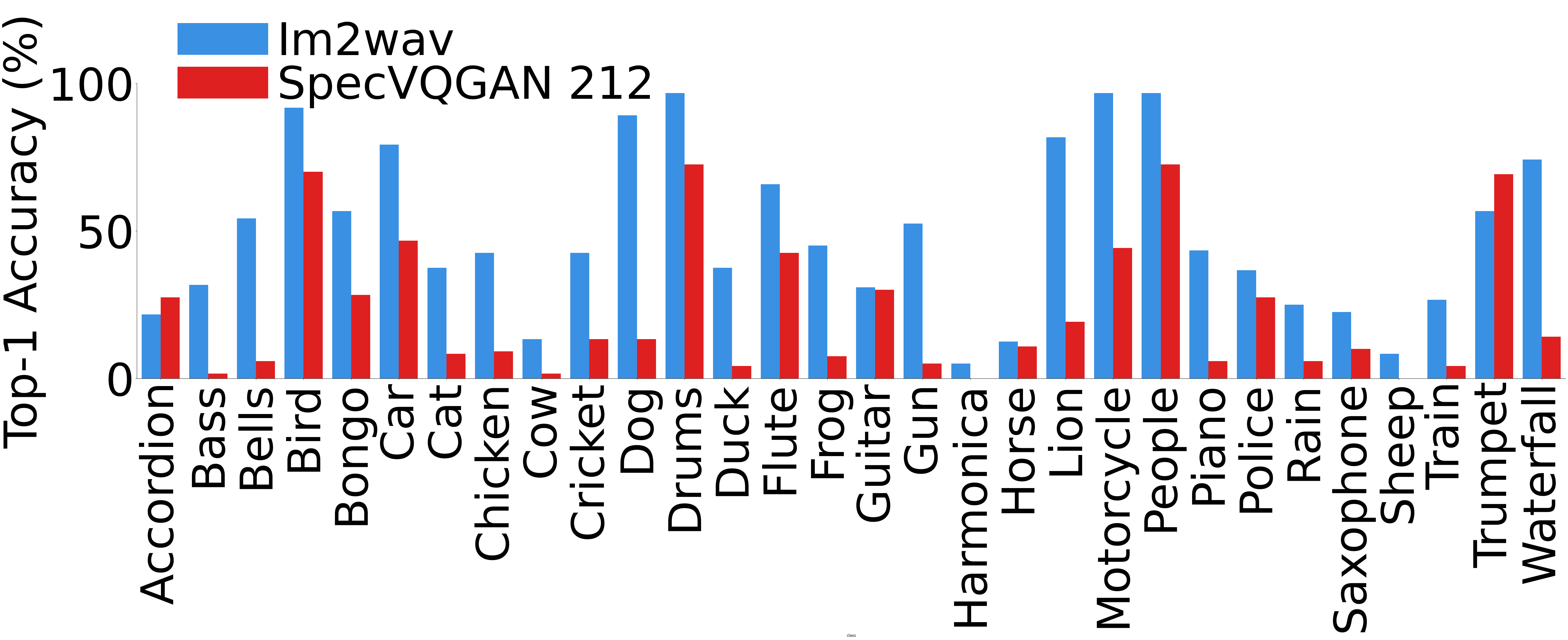}
         \caption{\label{fig:accimage}}
     \end{subfigure}
     \begin{subfigure}[b]{0.49\textwidth}
         \centering
         \includegraphics[width=\columnwidth]{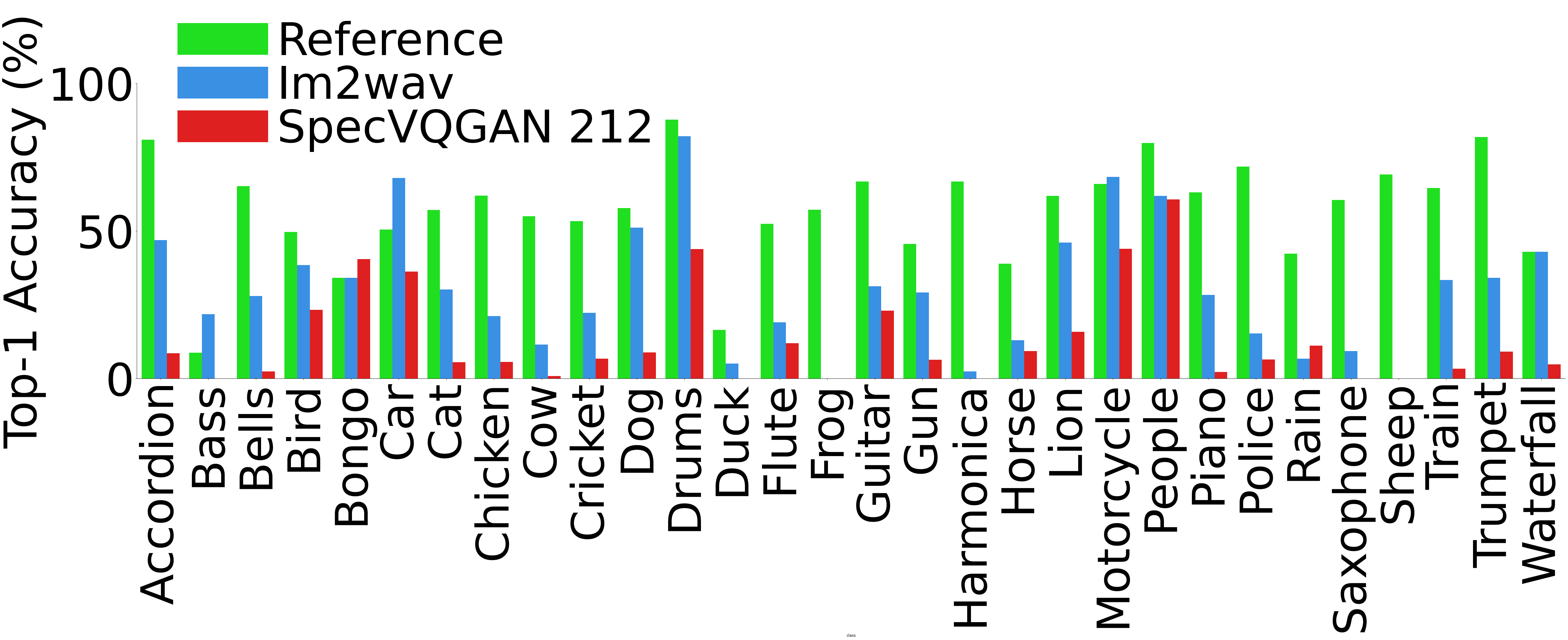}
         \caption{\label{fig:accvideo}}
     \end{subfigure}
    \caption{(a) Accuracy per class considering a single image condition from the \imhear{} dataset. (b) Accuracy per class considering a sequence of images (i.e., video) condition from the test-set of VGGSound dataset.\label{fig:acc}}
\end{figure*}

\vspace{-0.2cm}
\section{Experiments}
\label{sec:exps}
\vspace{-0.1cm}

\subsection{Experimental Setup}
\label{sec:setup}
{\noindent \bf Hyper-parameters.}
In all experiments we evaluate 4 seconds of generated audio, sampled at 16kHz. For the VQ-VAE model, we use a total of 5 convolutional layers with stride 2 for the encoder and the reversed operations for the decoder. The first codebook is applied after three convolutional layers, resulting in a downsampling factor of 8. Then, the second codebook is applied after two additional convolutional layers, resulting in an overall downsampling factor of 32. This corresponds to 2000 tokens per second in the \textsc{Up} model and 500 tokens per second in the \textsc{Low} model. Each codebook contains 2048 codes with embedding size of 128. For the auto-regressive models, we use a Transformer architecture with 48 layers and sparse attention, using a hidden size of 1024 dimensional vectors. For the CLIP model we use the $ViT-B/32$ version. Code is publicly available.

{\noindent \bf Data.}
We use the VGGSound dataset~\cite{Vggsound} extracted from videos uploaded to YouTube with audio-visual correspondence, containing $\sim$200k 10-second videos. We follow the original VGGSound train/test splits. For the evaluation, every test-set video is used with its initial 4 seconds only. To better evaluate the performance of the proposed method on out-of-distribution samples, we additionally collected 100 images from the web, containing 30 visual classes ($\sim$2-8 images per class), denoted as \imhear, and evaluate our method on it. To ensure our results are statistically significant we generate 120 audios per image class with each image used for an equal number of samples. This dataset is publicly available to support reproducibility and evaluation in future research.

\begin{table}[t!]
\centering
\caption{Main results: left part videos - VGGSound test-set \cite{Vggsound}, right part single image - \imhear{}.}
\label{tab:main}
\resizebox{\columnwidth}{!}{
\begin{tabular}{lcccc|cc}
\toprule
Method & FAD$\downarrow$ & KL$\downarrow$  & CS$\uparrow$ & ACC $\uparrow$ & CS$\uparrow$ & ACC $\uparrow$\\
\midrule                                       
Reference                              &  -    &  -    & \bf 7.61 & \bf 56.93\% & -    & -\\
\midrule
\cite{SpecVQGAN_Iashin_2021}~1 Feats   &  6.99 &  3.19 & 4.41 & 12.79\% & 5.54 & 21.92\%\\
\cite{SpecVQGAN_Iashin_2021}~5 Feats   &  6.81 &  3.13 & 4.54 & 14.44\% & 5.54 & 22.03\%\\
\cite{SpecVQGAN_Iashin_2021}~212 Feats &  6.64 &  3.10 & 4.62 & 14.44\% & 5.90 & 22.36\%\\
Ours                                   &  \bf 6.41 &  \bf 2.54 & \bf 7.19 & \bf 35.77\% & \bf 9.53 & \bf 49.14\%\\
\bottomrule
\end{tabular}}
\vspace{-0.1cm}
\end{table}

{\noindent \bf Evaluated Baselines.} We compare the proposed method to SpecVQGAN~\cite{SpecVQGAN_Iashin_2021}, a state-of-the-art open-domain visually guided audio generation model. We use the pre-trained models provided by the authors, using three ResNet50 Features-based models which were also trained on VGGSound. The difference between the three SpecVQGAN models is the required length of their conditioning image sequence.
SpecVQGAN operates in 21.5 fps. Therefore, when conditioning the 212 Feats model on 4-second videos, we repeat the last frame in order to reach its required number of frames. The same is done for the 212/5 Feats models when considering single image conditioning.

{\noindent \bf Evaluation Functions.} We evaluate the generated sounds on two aspects, fidelity (FAD) and relevance to the visual condition (KL, Accuracy and Clip-score). 

Adapting the Fr\'echet Inception Distance (FID) metric used to evaluate generative image models fidelity~\cite{fid} to the audio domain, Kilgour et al.~\cite{fad} proposed Fr\'echet Audio Distance (FAD). FAD measures the distance between the generated and real distributions. Features are extracted from both the real and generated data using an audio classifier~\cite{vggish} which was pre-trained on AudioSet~\cite{gemmeke2017audio}. The distributions of the real and generated extracted features are modeled as a multi-variate normal distributions $\mathcal{N}(\mathbf{\mu_{r}}, \mathbf{\Sigma_{r}}), \mathcal{N}(\mathbf{\mu_{g}}, \mathbf{\Sigma_{g})}$, respectively. 
The FAD is then given by the Fréchet distance between these distributions,
\begin{equation}
\label{eq:fad}
FAD = \twonorm{\mathbf{\mu_{r}} - \mathbf{\mu_{g}}}+\mathrm{tr}(\mathbf{\Sigma_{r}} + \mathbf{\Sigma_{g}}) -2\sqrt{\mathbf{\Sigma_{r}}\mathbf{\Sigma_{g}}}.
\end{equation}

Next, we adapt Clip-Score (CS), which has shown to be highly effective in evaluating image-caption correspondence~\cite{hessel2021clipscore, nichol2021glide}. We replace the CLIP text encoder with Wav2Clip model~\cite{wu2022wav2clip}, which is an audio encoder trained using contrastive loss on corresponding images and audio on top of the frozen CLIP image encoder. We pass both the image and the generated sound through their respective feature extractors. Then, we compute the expectation of cosine similarity of the resultant feature vectors, multiplied by a scaling factor, $\gamma$. We use $\gamma=100$ as in~\cite{nichol2021glide}.

Since a video is a sequence of images, each image CS is independently calculated with the whole audio and then we average the resulting CS. When dealing with longer or semantically complicated videos, we would consider applying this metric on short windows where one can expect a semantic shared by all the images and the audio. When experimenting with replacing the average with median, we observed similar results. This might indicate that a single averaged window is suitable for 4-second VGGSound test-set videos.

Lastly, we use PaSST~\cite{passt} audio classifier trained on AudioSet~\cite{gemmeke2017audio} to obtain a distribution over 527 classes. On top of the classifier output, we compute KL Divergence between the class distribution of the original samples and the generated ones. As we do not have the reference audio for the \imhear{} dataset, we also compute the accuracy of the classifier on the generated audio samples. For completeness, we report the accuracy also for VGGSound considering the \imhear{} classes only.

\begin{table}[t!]
\centering
\caption{Ablation Study: left part videos - VGGSound test-set \cite{Vggsound}, right part single image - \imhear{}.}
\label{tab:abbl}
\resizebox{\columnwidth}{!}{
\begin{tabular}{ccc|cccc|cc}
\toprule
 CFG & \textsc{Up} & \textsc{Every} & FAD$\downarrow$ & KL$\downarrow$ & CS$\uparrow$ & ACC$\uparrow$ & CS$\uparrow$ & ACC $\uparrow$\\
\midrule                               
\ding{55} &  \ding{55}   & \ding{55} & 12.47 & 3.05 & 5.16 & 19.68\% & 7.30 & 26.67\%\\ 
\ding{55} &  \ding{55}   & \ding{51} & 12.44 & 3.04 & 5.08 & 18.37\% & 7.23 & 23.56\%\\ 
\ding{51} &  \ding{55}   & \ding{55} & 10.23 & 2.76 & 5.94 & 30.07\% & 8.64 & 39.94\%\\ 
\ding{51} &  \ding{55}   & \ding{51} & 10.13 & 2.72 & 5.99 & 29.81\% & 8.87 & 41.97\%\\ 
\ding{55} &  \ding{51}   & \ding{55} & 8.86  & 2.85 & 5.89 & 22.51\% & 7.58 & 30.64\%\\ 
\ding{55} &  \ding{51}   & \ding{51} & 8.99  & 2.85 & 5.82 & 21.37\% & 7.55 & 29.61\%\\ 
\ding{51} &  \ding{51}   & \ding{55} &  6.52 & 2.58 & 7.12 & 35.01\% & 9.27 & 46.78\%\\ 
\midrule
\ding{51} &  \ding{51}   & \ding{51} & \bf 6.41  & \bf 2.54 &  \bf 7.19 & \bf 35.77\% & \bf 9.53 &  \bf 49.14\%\\             
\bottomrule
\end{tabular}}
\end{table}

\vspace{-0.1cm}
\subsection{Results}
\label{sec:results}

Table~\ref{tab:main}~summarizes the results for the proposed method and evaluated baselines. Results suggest that the proposed method is superior to the evaluated baselines both in terms of fidelity and relevance. 
All the evaluated models produce better relevance metrics when conditioned on \imhear{} single images than on VGGSound videos. The different relevance metrics keep the same ranking across the evaluated models. Figure~\ref{fig:acc} show that our model is capable of producing diverse sounds of more classes compared to SpecVQGAN~\cite{SpecVQGAN_Iashin_2021}.

\vspace{-0.2cm}
\subsection{Ablation study}
Next, we conduct an ablation study to better understand the effect of the different components of our proposed method summarized in Table~\ref{tab:abbl}. Specifically, we evaluate the effect of CFG, using the \textsc{Up} model, and conditioning on the temporally-corresponding frame at every token, denoted as \textsc{Every}.
Results suggest that \textsc{Up} has a noticeable effect on fidelity as all models with \textsc{Up} achieve lower FAD than all models without it regardless of CFG or \textsc{Every} usage.
Results suggest that CFG has a noticeable effect on visual relevance as all models with CFG achieve better relevance metrics than all models without it regardless of \textsc{Up} or \textsc{Every} usage. This fits the notion that the \textsc{Low} level learns the highest degree of abstraction, including the semantics, while the higher resolution \textsc{Up} level refines the abstract foundation to a more natural sound. Finally, the results suggest that \textsc{Every} has a relatively small effect on both fidelity and relevance metrics and improves when combined with CFG.
\vspace{-0.2cm}
\section{Conclusion \& Future Work}
\label{sec:con}
\vspace{-0.1cm}
In this work, we proposed \imwav{}, a method for open-domain image-to-audio generation. We empirically demonstrated that \imwav{} is superior to the evaluated baselines considering both fidelity and relevance metrics. In addition, we proposed \imhear{}, an out-of-domain benchmark for future evaluation of image-to-audio models. We hope that with the provided benchmark, comparing image-to-audio generation models will be standardized and more accessible. For future work, we would like to explore the effect of using multi-modal features for audio generation. Furthermore, we would like to evaluate \imwav{} considering complex multi-scene and multi-object image sequences.

\section*{Acknowledgements}
We would like to acknowledge support for this research from the Israeli Science Foundation (ISF grant 2049/22).

\bibliographystyle{IEEEbib}
\bibliography{refs}

\end{document}